\documentclass[prl,twocolumn,aps,superscriptaddress,showpacs,floatfix]{revtex4}
\usepackage{amsmath}
\usepackage{graphicx}
\usepackage{color}
\usepackage{epstopdf}

\begin{document}
\title{Fundamental energy limits in the physics of small-scale binary switches}
\author{L. Gammaitoni\cite{lgemail}}
\address{NiPS Laboratory, Dipartimento di Fisica, Universit\'a di Perugia, and Instituto Nazionale di Fisica Nucleare, Sezione di Perugia, I-06100 Perugia, Italy}
\author{D. Chiuchi\'u}
\address{NiPS Laboratory, Dipartimento di Fisica, Universit\'a di Perugia}
\author{M. Madami}
\author{G. Carlotti}
\address{Dipartimento di Fisica, Universit\'a di Perugia}

\date{\today}

\begin{abstract}
Binary switches are the basic element of modern digital computers. In this paper we discuss the role of switching procedure with reference to the fundamental limits in minimum energy dissipation. We show that the minimum energy depends on the switching procedure and test this result with micromagnetic simulations of a nanoscale switch realized with single cylindrical element of permalloy (NiFe). Finally we establish a relation between minimum energy and switching error probability.

\end{abstract}
 \maketitle

Is it possible to operate a computing device with zero energy expenditure? This question has been addressed during the second half of last century by a number of scientists and has led to a version of the second principle of thermodynamics that, assuming the Shannon information as a special form of the Gibbs-Boltzmann entropy, establishes that necessary condition to operate a computing device with zero energy dissipated is that the computing process does not decrease information\cite{1,2,3,4}. This result, often invoked as "Landauer principle", has been recently put under experimental test\cite{NotreDame,5,6} with the aim of exploring the limits in low power computation\cite{Bokor}.

On the other hand, it is a well-known fact that in the last forty years the semiconductor industry has been driven by its ability to scale down the size of the CMOS-FET\cite{7} switches, the building block of present computing devices, and to increase computing capability density up to a point where the power dissipated in heat during computation has become a serious limitation\cite{8,9}. In order to overcome such a limitation, since 2004 the Nanoelectronics Research Initiative\cite{12} has launched a grand challenge to address the fundamental limits of the physics of switches. In Europe the European Commission has recently funded a set of projects with the aim of minimizing the energy consumption of computing\cite{MINECC}. 
Notwithstanding a general agreement on the Landauer principle that would set at zero the energy required by a simple switching event (considered an information preserving transformation), among the researchers involved in designing the future ICT is widespread the idea that there is lower limit of $3 k_B T$ (approx $10^{-21} J$ at room temperature)\cite{pop}. Such a limitation would arise as a consequence of the switching procedure itself. In this paper we discuss the physics of the switching event and show that such a limit is not of fundamental nature and there is an optimal procedure to operate a generic switch according with the zero dissipation goal (we call this procedure \emph{zero-power switching}). 

\begin{figure}[b]
\includegraphics*[width=8.9cm]{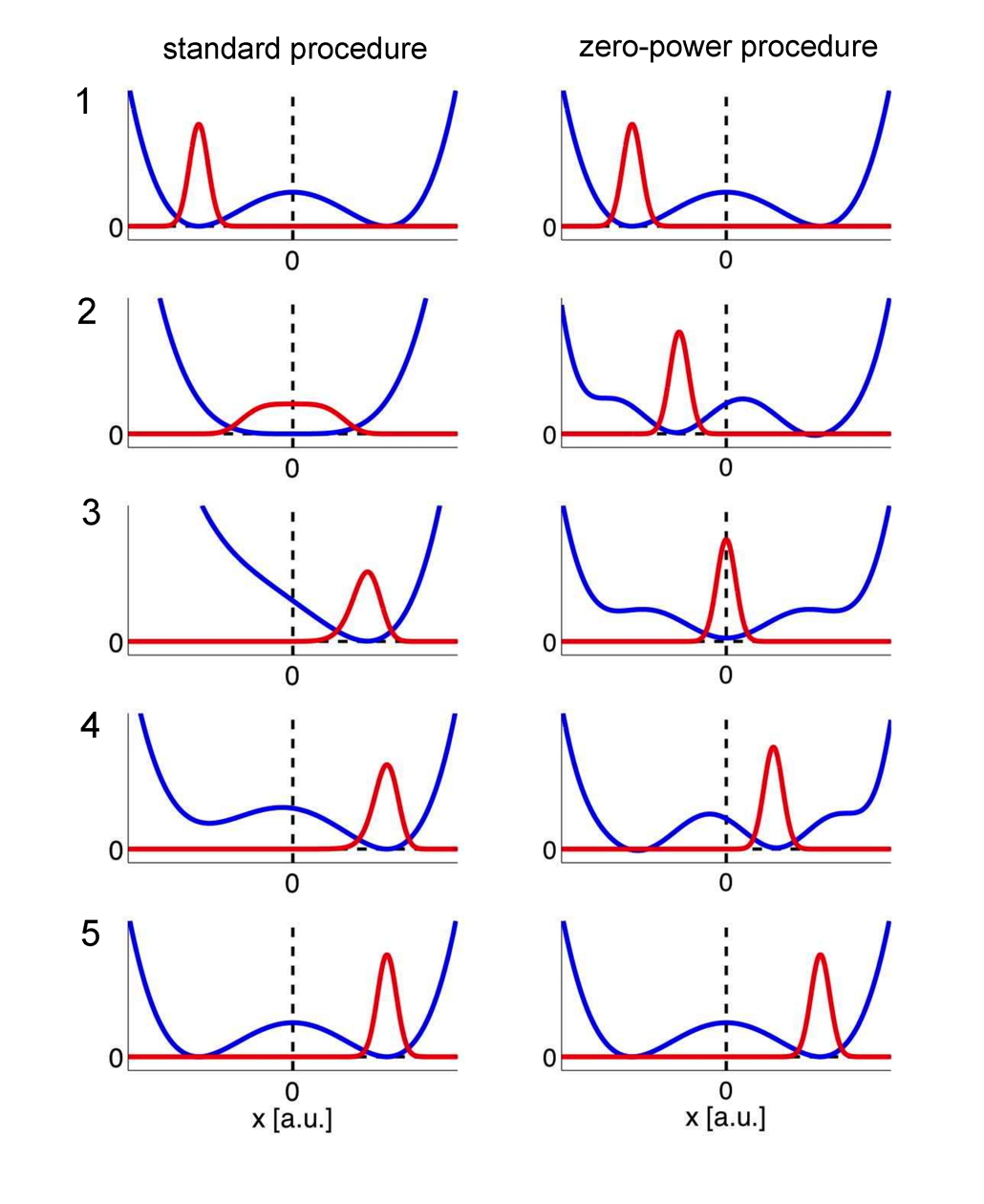}
\caption{
Schematic representation of a switching procedure. Left: $\emph{standard procedure}$. Changes in the probability distribution (red) and in the potential $U(x)$ (blu) due to the application of the external force $f(t)$. Step 1 to 5 from top to bottom.  Right:  $\emph{zero-power procedure}$. In order to satisfy condition $1)$ and $2)$ we apply slowly a proper external force $f(t)$ that keeps the average position of the particle always close to the minimum of the potential well and that does not change the probability density along the path (cond. $3)$, in a constant entropy transformation condition.
}
\label{F1}
\end{figure}

For what is relevant in this paper, the physics of a generic small-scale binary switch will be described in terms of a generic two-state system in contact with a thermal bath at constant temperature $T$. We discuss the switch dynamics in terms of the \emph{continuous dynamics of a single degree of freedom} $x(t)$, ideally representing a relevant observable of the switch dynamics, e.g. the position of a cursor, the quantity of electric charge, the value of magnetic or electric field, etc. In order to represent the binary character of the switch we need to identify the two logic states. We assume that the logic state $0$ ($1$) is associated with $x<0$ ($x>0$). The two states, in order to be dynamically stable, are separated by a potential energy barrier.

The continuous dynamics of $x(t)$ can be mathematically described by a proper Langevin equation\cite{Gardiner}:

\begin{equation}
\label{motion}
m  \ddot{x} = -{{dU(x)}\over{dx}} - \gamma \dot{x} + \xi(t) + f(x,t)
\end{equation}

Where $U(x)$ is a symmetric bistable potential and $f(x,t)$ is an external force that can be applied when we want to change logic state. $F(x,t) = -dU(x)/dx + f(x,t)$ is the total conservative force acting on the system. $\xi(t)$ represents the fluctuating force whose statistical features are connected with the dissipative properties $\gamma$ by a proper fluctuation-dissipation relation. 
The time evolution of the corresponding probability density $p(x,t)$ is usually described in terms of the associated Fokker-Planck equation\cite{Gardiner}. 
The two states: 0 and 1 are realized with probability respectively $p_0$ and $p_1$ ($p_0+p_1=1$) given by:

\begin{equation}
\label{prob}
p_0 = \int_{-\infty}^{0} {p(x,t) dx}, \qquad   p_1 = \int_{0}^{+\infty}{p(x,t) dx}
\end{equation}
 
In a generic computing device the practical switch is supposed to rest in one of the two logic states between two subsequent switching procedures. In our bistable model, due to the presence of the fluctuating force the particle will oscillate around the potential minima, with occasional random crossings of the potential barrier between the two wells. 
Symmetric potential and zero average fluctuating force implies at equilibrium $p_0 = p_1$.

State $0$ and $1$ have the same energy thanks to potential symmetry so, in order to perform the switch operation with zero energy expenditure, we need to reduce to zero the work performed by the forces acting on the system. The forces are represented here by the conservative force $F(x,t)$ and the dissipative force $\gamma \dot{x}$.  In the following we exclude the case of a reversible transformation where the work performed against the conservative forces during the switch is stored as potential energy in some external place and recovered subsequently because this condition is practically difficult to be implemented in a small scale switch. To begin our analysis let's start with the following procedure  (see Fig. 1 left, \emph{standard procedure})\cite{5}, that presents a switching from $0$ to $1$ (the $1$ to $0$ is analogous). We start with the system in the logic state $0$ (step 1 - first picture from the top). Here the potential barrier is initially lowered down to zero (step 2) and subsequently the potential is tilted in order to move the particle to the right well (step 3). Finally the barrier is raised to its initial value (step 4) and the tilt is removed (step 5). All five steps in this procedure can be performed by a proper application of the external force $f(x,t)$ in order to keep the average velocity of the particle close to zero (zero friction) and the average position in correspondence to an approximately zero derivative potential value (zero deterministic force). However, in spite of a negligible friction and negligible external work performed on the system, this procedure does not allow for a zero-power switching. This is due to the unavoidable entropy reduction occurring in steps 3-5, after the sudden increase in step 1-2, due to the barrier drop. This is apparent by the change in the probability distribution (see Fig. 1 left) and can be demonstrated quantitatively as follows. 
The system entropy in the various steps can be computed according to Gibbs as:

\begin{equation}
\label{Gibbs-e}
S_s = -k_B \sum_i{p_i \ln{p_i}}
\end{equation}
 
where we assume that the Gibbs entropy and the Shannon entropy coincide\cite{4}. $s$ indicates the step number and $i=0,1$. By the moment that in step 1 we have $p_0=1$ and $p_1=0$, $S_1 = 0$. In step 2 $p_0=p_1=1/2$ and thus $S_2 = k_B \ln(2)$.  Thus the change in entropy is $\Delta S_{2-1} = k_B \ln(2) > 0$. On the other hand from step 3 to step 5 entropy is reduced from $S_2$ to $S_5=S_1=0$, thus $\Delta S_{5-2} = - k_B \ln(2) < 0$. According to the thermodynamics, while the entropy increase can be performed without energy exchange, these last steps cannot be performed without energy expenditure. This result is also in agreement with what has been observed in the simulation of a nano-magnetic system\cite{Bokor} where this switch procedure has been applied.

Thus, based on this discussion, the conditions required to realize a zero-power switch can be summarized as: $1)$ The application of the external force keeps the average position of the particle always close to the minimum of the potential well (zero total force). $2)$ The switch event has to proceed with a speed as small as possible (zero friction approximation). $3)$ The system entropy must remain constant during the switch event. Fig.1 right, shows as an example, a possible procedure that satisfies these three conditions. 

\begin{figure}[b]
\includegraphics*[width=8.5cm]{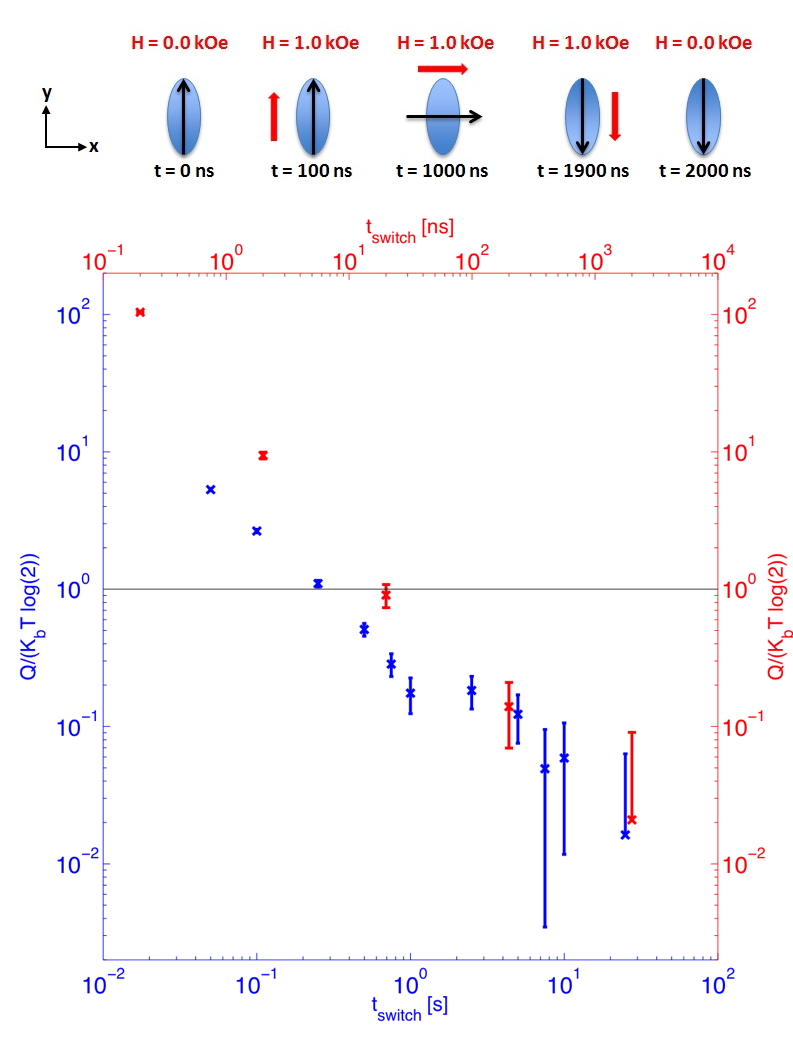}
\caption{Upper: five steps of the $\emph{zero-power}$ procedure (process A) applied to the magnetic dot. Lower: energy dissipated during the zero-power switching procedure $\emph{vs}$ overall switch time. Results from micro magnetic simulations (red) and  eq. (1) (blue). Micromagnetic simulations have been realized with a customized version of the commercial software Micromagus. Magnetic parameters have standard values for bulk permalloy: saturation magnetization $M_S = 800 \cdot10^3$ $A/m$, exchange stiffness constant $A = 1.3\cdot10^{-11}$ $J/m$, damping coefficient $\alpha = 0.01$ and $K_1 = 1.6 \cdot10^4$ $J/m^3$. 
}
\label{F2}
\end{figure}

To test these conclusions we performed micromagnetic simulations of a nano-scale switch, consisting of a single cylindrical element of permalloy (NiFe) with dimensions
$50\times 50\times 5$ $nm^3$ ,discretized in elementary cells of $5\times 5\times 5$ $nm^3$. All simulations have been performed at $T=300$ K  and the effect of thermal fluctuations is introduced into the simulation with a random fluctuating magnetic field which is delta correlated both in time and in space, with an amplitude which fulfills the fluctuation-dissipation theorem. 

In order to simulate a bistable system we introduced a uniaxial anisotropy in the plane of the NiFe dot, along the $y$ direction. This anisotropy defines two low energy states for the magnetization, $+y$ and $-y$, which can be defined as $0$ and $1$ states respectively. These two energy minima are separated by an energy barrier whose height is determined by the value of the uniaxial anisotropy constant $K_1$. 

The switching procedure was realized applying a sequence of external magnetic fields which is illustrated in the upper part of Fig.2. The presence of the uniaxial anisotropy along the $y$ direction has been represented in the figure using an elliptical shape with its major axis parallel to $y$. The system starts within no applied external field and with the initial magnetization oriented along the positive $y$ direction (i.e. the system is in one of the two energy minima). In the first stage a positive external field ($H$) is applied along $y$ with a slope, up to a maximum value of $H=1.0$ $kOe$. After that the external field is slowly rotated in the $x-y$ plane of $180^o$ towards the negative $y$ direction, keeping its amplitude fixed at the value $H=1.0$ $kOe$. During the final stage of the simulation the external field is removed with a negative slope, opposite to that of the first stage. The energy dissipated during the switching procedure is then calculated as the integral of the scalar product $HádM$ where $M$ is the average magnetization of the system. The total switch time was also varied in the range $2á(10^{-1}-10^3)$ $ns$. The results shown in Fig.2 (lower) for micromagnetic simulations represents the energy dissipation during the switching procedure as a function of the total switch time. The calculations were performed for a barrier height of about $40$ $k_BT$. In this switching procedure the external field is applied parallel to the initial magnetization configuration so the system remains in the absolute minimum energy state during the whole procedure. In the same figure are also shown, for comparison, the results obtained with numerical solution of eq. (1)\cite{seifert}. Both sets of data clearly shows that the dissipated energy can be made as small as needed just increasing the total switch procedure duration, asymptotically reaching the zero energy dissipation limit, according to our prediction.
Based on these considerations for the zero-power switching we can address critically the recent observation\cite{NotreDame} that in order to reach zero-power switch operation, a key element is represented by the existence of a copy of the switch already in the final destination status. We have shown here that this condition is not required in a binary switch.

Finally we derive a relation that connects the minimum energy and the switch error probability.
Provided that a procedure for zero-power switching $0-to-1$ ($1-to-0$) exists, if this procedure, when applied to the initial state $1$ ($0$) produces no change in the state, i.e. the final state is still the state $1$ ($0$), then its minimum energy cost is $2 k_B T \ln(2)$.
	This is easily demonstrated by assuming that we want to use this procedure for resetting to $1$ (to $0$). We can compute the energy dissipated as the sum of two processes that are realized with probability $1/2$ each: the switch $0-to-1$ ($1-to-0$) with energy $Q_A$ and the transformation $1-to-1$ $(0-to-0)$ with energy $Q_B$. The total energy dissipated, due to the Landauer principle is $Q = 1/2\, Q_A + 1/2\, Q_B \geq k_B T \ln(2)$. However in hypothesis it was $Q_A=0$ thus $Q_B \geq 2 k_B T \ln(2)$. 
	
In Fig.3 we show the results of the digital simulation of eq. (1) together with the micromagnetic simulations for this case.
In this set of simulations the initial magnetization of the system was set to be antiparallel to the applied external field realizing process $B$. Here the system starts on a relative energy minimum which becomes more and more unstable, as the external magnetic field is increased in the opposite direction, until the magnetization is forced to reverse towards the absolute minimum energy state. Simulations show that the energy dissipated is much larger with respect to the process $A$ (in Fig. 2) and it can be reduced toward the limit $2 k_B T \ln(2)$ while reducing the energy barrier height.

\begin{figure}[b]
\includegraphics*[width=8.5cm]{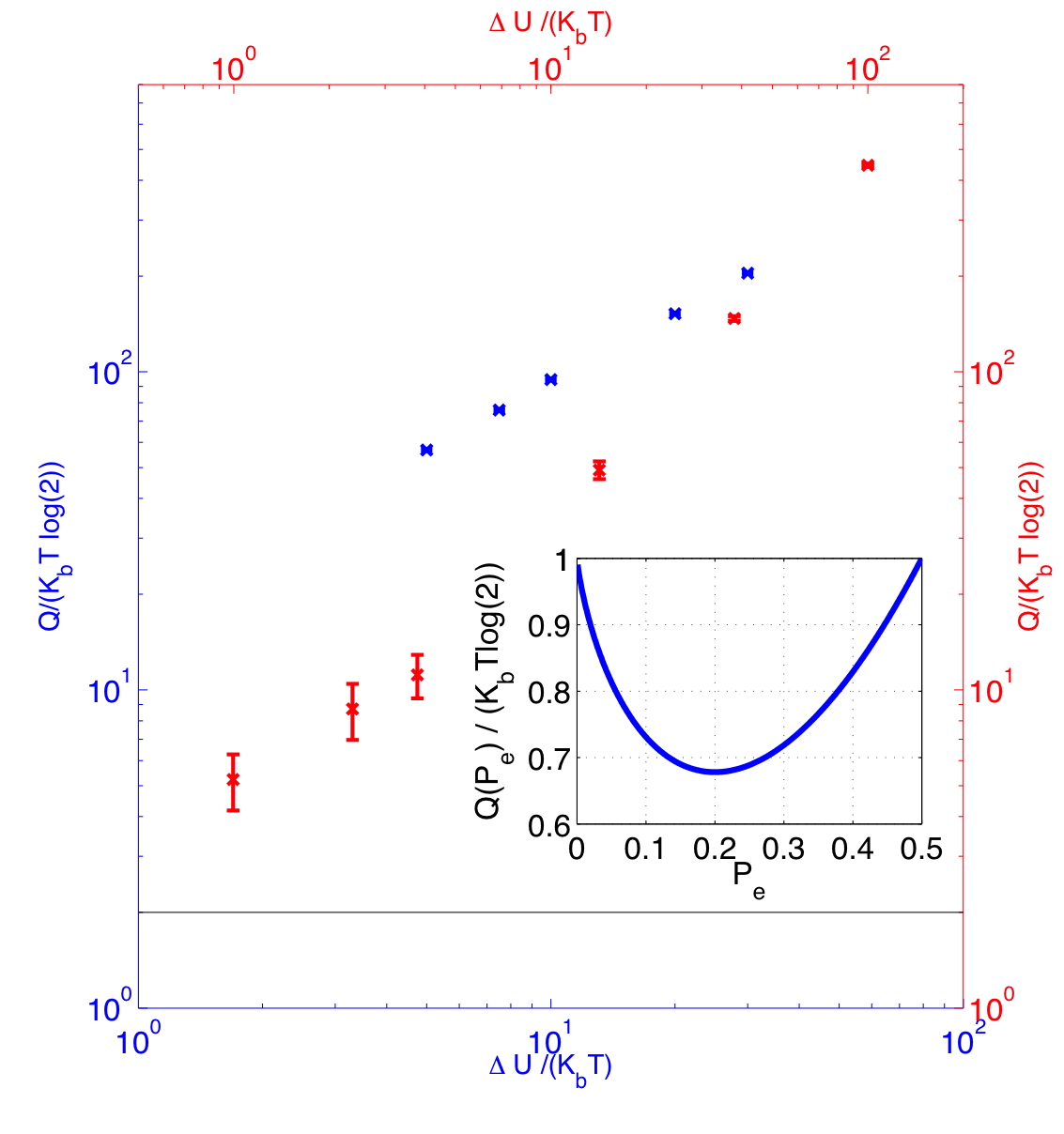}
\caption{Energy dissipated during the zero-power switching procedure, process $B$, $\emph{vs}$ barrier height. Results from micro magnetic simulations (red) and  eq. (1) (blue). The horizontal line represents the minimum energy toll at $2 k_B T \ln(2)$.
Inset: minimum energy $Q(P_e)$ for the reset-and-switch optimal procedure, as a function of the error probability $P_e$.}
\label{F3}
\end{figure}

This result has an interesting interpretation in terms of the minimum energy cost of a bit flip error. Bit flips may occur in a switch as a result of large fluctuations in the presence of a finite barrier height \cite{Gamma-speed}. As a consequence of a bit flip error, the switch is assumed to be in the wrong logical status, thus when we apply the proper zero-power procedure we incur in a minimum energy toll of $Q_{bf}=2 k_B T \ln(2)$, as just demonstrated. This argument is reminiscent of the minimum energy toll required when operating the electronic switches\cite{9} but at difference with that one, this is of a fundamental nature and applies regardless to the technology employed to build the switch. 
	
We observe that this result poses a limit to the trading between energy and uncertainty\cite{Luca} in the reset operation. As it was observed\cite{Luca}, the Landauer limit can be beaten by accepting a finite error probability $P_e$ during the reset. However, such a gain is now balanced by the additional energy dissipation due to the switch error condition that affects the subsequent switch.
By putting together the two contributions we can determine the minimum energy associated with the sequence reset-and-switch as a function of the reset error probability. The relation that generalize the one obtained in \cite{Luca} is expressed by: 
$Q(P_e) = k_B T ((1- P_e) \ln(1-P_e) + P_e \ln(P_e)) + k_B T \ln(2) + P_e 2 k_B T \ln(2)$.

In Fig. 3  (inset) we plot the minimum energy $Q(P_e)$ as a function of the error probability $P_e$. Remarkably $Q(P_e)$ shows a minimum $Q(\bar P_e) = k_B T \ln(8/5)$, for $\bar P_e = 1/5$ independent from $T$.
Such a behavior, characterized by the existence of an optimal error probability (e.g. due to a given noise intensity) that minimizes the reset-and-switch energy is reminiscent of a vast class of phenomena where the presence of a finite amount of fluctuations, instead of being detrimental, results beneficial \cite{SR}.

We thank C. Diamantini, I. Neri, G. Gubbiotti for useful discussions and acknowledge support by the European Union (FPVII (2007-2013) under G.A. n 318287 LANDAUER, G.A. n 270005 ZEROPOWER and G.A. n. 611004 ICT-Energy.


\begin{references}

\bibitem[*]{lgemail} e-mail: luca.gammaitoni@nipslab.org

\bibitem{1} J. von Neumann, Fourth University of Illinois lecture, in Theory of Self-Reproducing Automata, A.W. Burks, ed., p. 66. Univ. of Illinois Press, Urbana (1966).
\bibitem{2} R. Landauer, IBM J. Res. Dev. 5 (1961) 183-191.
\bibitem{3} C. H. Bennett, Int. J. Theoretical Physics 21 (1982) 905-940.
\bibitem{4} O. J. E. Maroney, Phys. Rev. E, 031105 (2009).
\bibitem{NotreDame} Alexei O. Orlov, Craig S. Lent, Cameron C. Thorpe, Graham P. Boechler, and Gregory L. Snider, Japanese Journal of Applied Physics 51 (2012).
\bibitem{5} A. Berut,	A. Arakelyan, A. Petrosyan, S. Ciliberto, R. Dillenschneider, and E. Lutz, Nature 483,187189 (2012).
\bibitem{6} S.Toyabe, T. Sagawa, M. Ueda, E. Muneyuki, and M. Sano, Nature Physics 6, 988992 (2010).
\bibitem{Bokor} B. Lambson, D. Carlton, and J. Bokor, Phys. Rev. Lett. 107, 010604 (2011).
\bibitem{7} CMOS (Complementary MetalOxideSemiconductor), FET (Field-Effect Transistor).
\bibitem{8} K. Bernstein, R.K. Cavin, W. Porod, A. Seabaugh, J. Welser, Proceedings of the IEEE , vol.98, no.12, pp.2169-2184, Dec. 2010. 
\bibitem{9} J. Welser, G.Bourianoff, V. Zhirnov and R. Cavin, J. Nanopar. Res., 10:1-10 (2008). 
\bibitem{12} The Nanoelectronics Research Initiative (nri.src.org) 
\bibitem{MINECC} FET Proactive Intiative: Minimising Energy Consumption of Computing to the Limit (MINECC). $http://cordis.europa.eu/fp7/ict/fet-proactive/minecc_en.html$
\bibitem{pop} E. Pop, Nano Res. (2010) 3: 147169.
\bibitem{Gardiner} C.W. Gardiner, Handbook of Stochastic Methods (Springer, Berlin). 1985.
\bibitem{Gamma-speed} L. Gammaitoni, Appl. Phys. Lett., 91, 224104, 2007.
\bibitem{seifert} U. Seifert, Rep. Prog. Phys. 75: 126001, 2012.
\bibitem{Luca} L. Gammaitoni, 2011, arXiv:1111.2937; L. Gammaitoni, Nanoenergy Letters, 5, 10, 2013.
\bibitem{SR} L. Gammaitoni, P. Hanggi, P. Jung, and F. Marchesoni, Rev. Mod. Phys. 70, 223 1998.



\end{references}
\end{document}